\documentclass{aa}  

\usepackage{bbm}
\usepackage{graphicx}
\usepackage[varg]{txfonts}
\usepackage{xcolor}
\usepackage[colorlinks=true, linkcolor=blue, urlcolor=blue, citecolor=blue]{hyperref}
\usepackage[capitalize]{cleveref}

\crefname{section}{Sect.}{Sects.}
\Crefname{section}{Section}{Sections}

\crefname{figure}{Fig.}{Figs.}
\Crefname{figure}{Figure}{Figures}

\crefname{equation}{Eq.}{Eqs.}
\Crefname{equation}{Equation}{Equations}

\begin{document}

\title{Gaussian process-based Bayesian imaging of bolometric (sub-)millimetre data}

\author{
   Richard~Fuchs \inst{1}
   \and
   Jonas~W\"urzinger \inst{1,2}
   \and
   Joshiwa~van~Marrewijk \inst{3}
   \and
   Thomas~W.~Morris \inst{4}
   \and
   Tony~Mroczkowski \inst{5,6}
   \and
   Luca~Di~Mascolo \inst{7}
   \and 
   Aileen~Chen \inst{1}
   \and
   Adriaan~J.~Duivenvoorden \inst{8}
   \and 
   Mark~Devlin\inst{9}
   \and 
   Simon~Dicker\inst{9}
   \and
   Lukas~Heinrich \inst{1,2}
}

\institute{
   Technical University of Munich, TUM School of Natural Sciences, Boltzmannstr. 2, 85748 Garching, Germany \\
   \email{richard.fuchs@tum.de}
   \and
   Excellence Cluster ORIGINS, Boltzmannstrasse 2, 85748 Garching, Germany
   \and
   Leiden Observatory, Leiden University, P.O. Box 9513, 2300 RA Leiden, The Netherlands
   \and
   Yale University, New Haven, CT 06511, USA
   \and
   Institute of Space Sciences (ICE-CSIC), Carrer de Can Magrans, s/n, 08193 Cerdanyola del Vall\`es, Barcelona, Spain
   \and
   Institut d'Estudis Espacials de Catalunya (IEEC), E-08860 Castelldefels, Barcelona, Spain
   \and 
   Kapteyn Astronomical Institute, University of Groningen, Landleven 12, NL-9747 AD, Groningen, The Netherlands
   \and 
   {Max-Planck-Institut für Astrophysik, Karl-Schwarzschild-Str. 1, 85748 Garching, Germany}
   \and
    University of Pennsylvania, 209 S. 33rd St., Philadelphia, PA 19014, USA
}


\date{Received XXXX; accepted XXXX}

\abstract{
Widefield single-dish telescopes operating at (sub-)mm wavelengths ($\lambda \sim 4$~mm to 350~$\mu$m) tend to feature much faster mapping speeds than radio interferometers at the same wavelengths, and are more suitable for measuring signals on larger spatial scales. However, the data are dominated by atmospheric fluctuations on large scales and by photon and instrumental noise on smaller scales, making it difficult to recover the true astronomical sky. We build on \texttt{maria-nifty}, a Gaussian process-based Bayesian imaging framework for reconstructing sky maps from single-dish telescope data, implemented on top of Numerical Information Field Theory (NIFTy). It uses modular generative models to efficiently separate the astronomical signal from atmospheric emission while providing pixel-wise uncertainty quantification for the reconstructed sky maps. Moreover, our method cleanly separates individual astrophysical signals, even though they are superposed on the sky. Following its successful validation on synthetic time-ordered data generated using the \texttt{maria} software, we apply it to real data observed with the MUSTANG-2 bolometric array on the 100-meter Green Bank Telescope. For the galaxy cluster RXJ1347.5-1145, we separate the extended cluster emission from the compact source at its centre and obtain a smoother map with fewer artefacts compared to a standard maximum-likelihood reconstruction using the same data. The JAX-based implementation and GPU acceleration provide the basis for applying the method to the vastly larger datasets that the next generation of single-dish telescopes will deliver.
}


\maketitle
\nolinenumbers

\section{Introduction} \label{sec:introduction}

Widefield sub-millimetre and mm-wave single-dish telescopes (at frequencies $\sim$90--950~GHz) feature faster mapping speeds and are more sensitive to large angular scales than their interferometric counterparts at the same wavelengths \citep{frayer2017, plunkett2023}. This makes them particularly well suited to observing extended emission, such as the thermal Sunyaev--Zeldovich (SZ) effect, which is a signal due to inverse Compton scattering of photons from the Cosmic Microwave Background (CMB) off hot ($\sim 10^8$ K) electrons in reservoirs of hot thermal plasma such as found in galaxy clusters (\citealt{Sunyaev1972}; see \citealp{mroczkowski2019} for a review). One way to probe the SZ effect is through large bolometric array detectors such as MUSTANG-2, a 223-detector array on the 100-meter Green Bank Telescope (GBT) that provides $\approx 9\arcsec$ angular resolution continuum mapping in a broad bandwidth centred around a frequency of 93~GHz (see \citealt{dicker2014,dicker2014b} for more detail).

However, ground-based bolometric observations are dominated by the emission of the atmosphere. Fluctuations of the water vapour along the line of sight cause a signal that follows a steep, Kolmogorov-like temporal power spectrum \citep{lay1997, errard2015} and varies on short timescales as the telescope scans across the sky \citep{morris2022, morris2025}. While the atmosphere dominates on large scales, where it exceeds the astronomical signal by several orders of magnitude, the data on smaller scales are limited by photon and instrumental noise \citep{hill2024}. Separating these contributions from the subdominant astronomical signal is the central challenge in recovering the true sky.

Traditional map-making methods remove the atmospheric contribution before the creation of the map. In doing so, a common mode or the leading principal components of the focal plane are subtracted from the observed data, which are then high-pass filtered and binned into a map \citep{perotto2020, romero2020}. This efficiently suppresses the atmosphere, but it also removes the astronomical signal on angular scales comparable to the size of the field of view (FOV) and beyond. It also requires a transfer function, which has to be calibrated with simulations. Maximum-likelihood map-makers reduce this loss by describing the atmosphere as correlated noise with an estimated covariance, which results in flatter transfer functions \citep{dunner2013, naess2025}. However, these are not very well constrained on large modes, which leads to large-scale gradients in the map, commonly referred to as ``bowling''. Examples may be found in \cite{romero2020} and \cite{orlowskischerer2022}, among others, who used the \texttt{minkasi}\footnote{\url{https://github.com/sievers/minkasi}} mapper to reconstruct the sky from MUSTANG-2 observational data.

At the same time, a new generation of single-dish facilities with much larger detector arrays and faster mapping speeds is emerging. These include the Simons Observatory \citep{solatscience2025}, the 6-meter Fred Young Submillimeter Telescope \citep[FYST;][]{ccat2023}, and the 50-meter Atacama Large Aperture Submillimeter Telescope (AtLAST; \citealp{klaassen2020, ramasawmy2022, mroczkowski2025, cicone2026}). Their key science cases, such as resolving the hot and ionized universe or the circumgalactic medium, rely on the reconstruction of exactly those large angular scales \citep{booth2024, lee2024, dimascolo2025}.

This requires realistic simulations, both to forecast their performance and to develop suitable map-making methods before their data are available. Simulators such as \texttt{TOAST} \citep{kisner2023} or \texttt{gateau} \citep{moerman2026} generate synthetic data for a given instrument and scanning strategy. In \texttt{maria}\footnote{\url{https://thomaswmorris.com/maria}} \citep{morris2022, marrewijk2024}, the simulated detector array scans through an evolving, turbulent atmosphere that is generated from satellite and ground-based weather measurements. The resulting data have been shown to be quantitatively similar to real MUSTANG-2 observations, which makes them a realistic benchmark for new map-making methods.

Apart from the simulations, the map-making itself must be improved for these facilities, which will deliver data from of order one hundred thousand detectors \citep{solatscience2025}. Information field theory provides a suitable framework for this task, casting the reconstruction of a physical field from incomplete and noisy data as a Bayesian inference problem \citep{ensslin2019}. Its numerical implementation \texttt{NIFTy}\footnote{\url{https://github.com/NIFTy-PPL/NIFTy}} \citep{selig2013, nifty5} combines Gaussian process (GP) priors with variational inference \citep{frank2021} and is written in \texttt{JAX}\footnote{\url{https://github.com/jax-ml/jax}} to scale to high-dimensional problems \citep{edenhofer2024}. It has been applied successfully to imaging problems across astrophysics, ranging from radio interferometry \citep{arras2022, roth2024, fuchs2026} and X-ray imaging \citep{westerkamp2024, eberle2024, guardiani2025} to $\gamma$-ray astronomy \citep{scheel2023}, strong lensing \citep{ruestig2024}, and three-dimensional dust reconstructions \citep{edenhofer2024dust}. In all of these applications, the signal, its correlation structure, and the instrumental effects are inferred jointly, together with their uncertainties.

\citet{wuerzinger2025} combined \texttt{maria} and \texttt{NIFTy} to recover the astronomical signal from simulated data. The implementation is under active development and publicly available on GitHub\footnote{\url{https://github.com/jwuerzinger/CMB_denoising}}. By introducing separate model components for map and atmosphere, we considerably improved the reconstructions compared to traditional map-making methods. In this work, we extend this framework to handle real bolometric single-dish data, marking the first application of its kind in sub-mm map-making.

In \Cref{sec:method}, we introduce the generative model, the likelihood, and the inference method of \texttt{maria-nifty}. In \cref{sec:applications}, we then apply it to MUSTANG-2 commissioning observations of RXJ1347.5-1145 (a.k.a.\ MACS J1347.5-1144, MCXC J1347.5-1144, or PSZ1 G324.05+48.79; hereafter referred to simply as RXJ1347) and compare it to a \texttt{minkasi} reconstruction of the same data.

\section{Methodology} \label{sec:method}

We cast the image reconstruction as a Bayesian inference problem, in which we seek to estimate the posterior distribution $P(s|d)$ of the sky map $s$ given the observed data $d$. The posterior is defined according to Bayes' theorem
\begin{equation} \label{eq:bayes}
   P(s|d) = \frac{P(d|s) P(s)}{P(d)},
\end{equation}
with the likelihood $P(d|s)$, the prior $P(s)$, and the evidence $P(d)$.

\subsection{Measurement equation} \label{sec:measurement-eq}

MUSTANG-2 is a bolometric array installed on the 100-meter Green Bank Telescope (GBT), located in Green Bank, West Virginia, USA. It operates at a frequency around 93 GHz and consists of 223 detectors, providing a field of view of approximately 4 arcminutes. The instrument observes the sky in the mm-wave regime, making it particularly sensitive to the thermal SZ effect from clusters such as RXJ1347 as well as thermal emission from cold ($\lesssim 100$~K) dust and gas in various astrophysical environments. For further details on the instrument, see \cite{dicker2014, dicker2014b}.

\begin{figure*}
\centering
   \includegraphics[width=17cm]{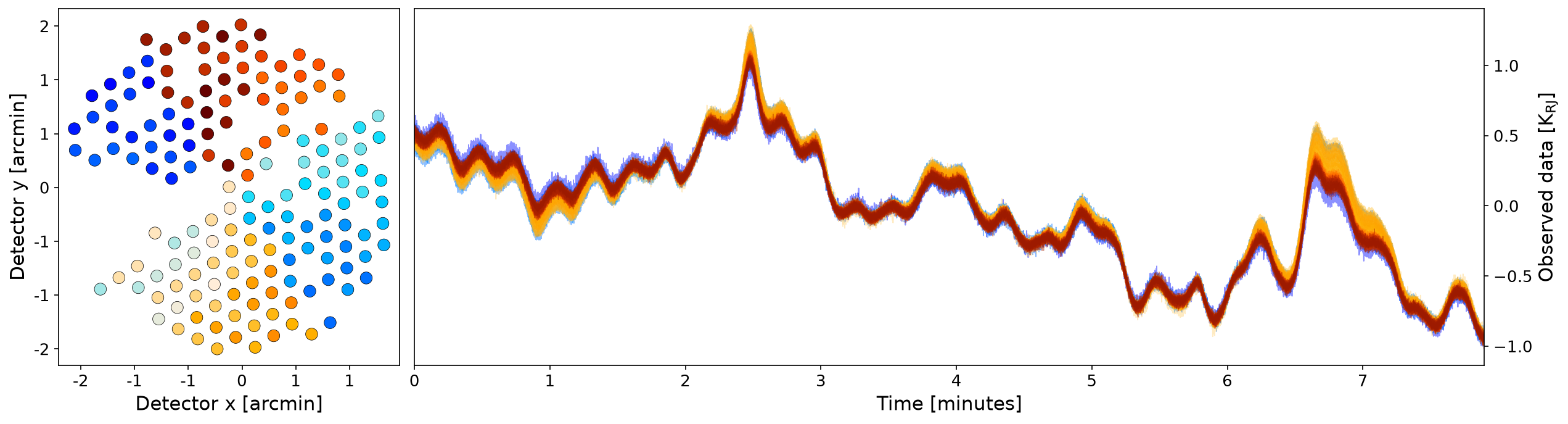}
     \caption{Time-ordered data for a single MUSTANG-2 scan of RXJ1347. \textit{Left:} positions of the 163 active detectors in the focal plane relative to the array centre. \textit{Right:} the observed TODs, showing the measured signal of each detector as a function of time step. In both panels, the detectors are coloured according to their index. The TODs are dominated by atmospheric emission, which manifests as large-scale fluctuations that are highly correlated across detectors, while the faint astronomical signal and instrumental noise are subdominant.}
     \label{fig:data}
\end{figure*}

The observations obtained from MUSTANG-2 are in the form of time-ordered data (TODs), which represent the signal measured by each detector as a function of time and position as the telescope scans across the sky. \Cref{fig:data} shows the detector positions as seen from the focal plane and the corresponding, mean-subtracted TODs for a single MUSTANG-2 scan of RXJ1347. 
Since the telescope is ground-based, the observed TODs are significantly affected by atmospheric emission, which can vary on short timescales and dominate the measured signal. Additionally, instrumental noise contributes to the observed data, further complicating the reconstruction of the true astronomical signal. The measurement equation for a single-dish bolometric array can be expressed as
\begin{equation} \label{eq:mse}
   d(t) = R(s, t) + a(t) + n,
\end{equation}
where $d(t)$ is the observed data at time $t$, $s$ is the sky map processed by the response function $R(s, t)$ of the given instrument, $a(t)$ represents the atmospheric emission, and $n$ is the instrumental noise. The response function first convolves the sky map with the beam profile, which we model as a Gaussian kernel whose full width at half maximum is given by the angular resolution of the instrument. Then, the convolved map is sampled at the positions corresponding to the scanning pattern of the telescope. For MUSTANG-2, the Lissajous daisy pattern is used, which provides good coverage of the field of view with higher sensitivity in the centre compared to the outer parts. In \texttt{maria-nifty}, we use a \texttt{JAX}-based version of the response function implemented in \texttt{maria}, which allows us to efficiently compute the response function and its gradients with respect to the model parameters, enabling gradient-based optimization during inference.

Strictly speaking, the noise term in \cref{eq:mse} is time-dependent as well, $n = n(t)$. As the telescope scans the sky, the statistical properties of the perceived noise change over the course of an observation \citep{dunner2013}. Many map-making methods, however, rely on stationary noise described by a single covariance estimated from the data. We account for this behaviour in our time-dependent atmospheric component $a(t)$ with individual model realizations for each detector, which capture the slowly varying contributions of both the atmosphere and the detector drifts, leaving only white noise in $n$.

\subsection{Prior models} \label{sec:prior-model}

The main challenge in reconstructing the sky map from the observed data lies in effectively separating the astronomical signal from the atmospheric contributions and instrumental noise. To achieve this, we construct a generative model for synthetic TODs
\begin{equation} \label{eq:gen_model}
   f(\xi) = R(s(\xi^s)) + a(\xi^a),
\end{equation}
that consists of a map $s(\xi^s)$ and an atmospheric component $a(\xi^a)$. Here, we drop the time dependence of $R$ and $a$. The parameters of the generative model are denoted by $\xi = \{\xi^s, \xi^a\}$, where $\xi^s$ and $\xi^a$ represent the parameters of the map and atmospheric models, respectively. Both model components are encoded as standardized generative models \citep{knollmuller2018}, which map a set of standard-normally distributed random variables $\{\xi\}$ to the desired target distribution. We can rewrite Bayes' theorem in a standardized form as
\begin{equation} \label{eq:bayes_std}
   p(\xi|d) = \frac{p(d|\xi) p(\xi)}{P(d)},
\end{equation}
where $p(\xi)$ is the prior distribution of the generative model parameters, which are standard-normally distributed by construction. Therefore, all prior assumptions about the map and atmosphere are encoded in the likelihood term $p(d|\xi)$ via the generative model $f(\xi)$.

\subsubsection{Map prior} \label{sec:map-prior}

The astronomical signal is assumed to be static during the time of an observation. Moreover, the observed sky map $s$ can consist of one or multiple components,
\begin{equation} \label{eq:map_components}
   s(\xi^s) = \sum_i s^i(\xi^{s_i}),
\end{equation}
where each component $s^i$ with model parameters $\xi^{s_i}$ can be defined on a different FOV, allowing us to capture different spatial scales and structures in the astronomical signal.

In general, most astronomical sources are expected to be smooth and spatially correlated. Therefore, we describe each map component as a two-dimensional correlated field
\begin{equation} \label{eq:gp}
   s^i \sim \mathcal{G}(s^i, S),
\end{equation}   
where $S$ is the covariance function that encodes the spatial correlation structure of the component $s^i$. Since the exact form of the correlation structure is unknown a priori, we use a non-parametric model to describe the covariance function, allowing us to infer it jointly with the signal. It is implemented in the correlated field model (CFM) of \texttt{NIFTy} in terms of a generative process
\begin{equation} \label{eq:cfm}
   c^i = A\xi^{c_i} ~~\text{with}~~ S = A A^\dagger.
\end{equation}
Assuming homogeneity and isotropy, the covariance function $S$ can be defined via a one-dimensional power spectrum, which describes the variance of the signal as a function of spatial frequency. For more details on the CFM and its implementation, we refer to \cite{arras2022}. 

For compact sources, whose emission is known to be positive, we additionally enforce positivity by passing the correlated field through a softplus function,
\begin{equation} \label{eq:softplus}
   c^i = \varphi\bigl(A\xi^{c_i}\bigr) ~~\text{with}~~ \varphi(z) = \log\bigl(1 + e^z\bigr),
\end{equation}
where $\varphi$ acts on each pixel separately. In contrast to an exponential, the softplus grows only linearly for large arguments, which avoids extreme prior samples, while it is smooth and differentiable everywhere and thus well suited for the gradient-based inference. Since the SZ signal appears as a decrement at the observed frequency, restricting the compact components to positive values keeps them from absorbing cluster emission and thereby improves the separation of the different superposed signals.

To allow the prior to modulate the statistically homogeneous and isotropic correlated field by a spatially varying amplitude, we multiply the output of the correlated field with a two-dimensional Gaussian with learnable widths \citep{ruestig2024}
\begin{equation} \label{eq:map_window}
   g^i(x) = \exp\left(-\frac{1}{2}\sum_d\frac{x_d^2}{\bigl(\sigma^i_d(\xi^{g_i})\bigr)^2}\right),
\end{equation}
where $x_d$ is the pixel coordinate along the direction $d$ relative to the centre of the corresponding FOV and $\sigma^i_d$ the width of the Gaussian in the same direction, given in units of the FOV and generated from the standard-normally distributed parameters $\xi^{g_i}$. A map component is then given by the element-wise product
\begin{equation} \label{eq:map_full}
   s^i(\xi^{s_i}) = g^i \odot c^i ~~\text{with}~~ \xi^{s_i} = \{\xi^{c_i}, \xi^{g_i}\},
\end{equation}
so that the widths of the Gaussian are inferred jointly with the correlated field. This ensures that the reconstructed map falls off to zero at the edges of the FOV, which is a reasonable assumption for many astronomical sources. Moreover, it breaks degeneracies between the map and atmospheric contributions and helps to avoid bowling in the map.

\subsubsection{Atmospheric prior} \label{sec:atm-prior}

The atmospheric emission is primarily dominated by fluctuations in water vapour and can vary on short timescales. In general, the atmospheric component can be modelled using a one-dimensional version of the CFM for each detector \citep{wuerzinger2025}. For the application to real data with an increasing number of TODs, however, the CFM turns out to have too many degrees of freedom, leading to degeneracies between the map and atmospheric contributions and making the inference problem more difficult to solve.

To circumvent this issue, we now use a simpler, parametric model with fewer degrees of freedom for the atmospheric emission. We model the atmospheric emission for each detector $i$ separately as a linear combination of $n$ B-spline basis functions,
\begin{align}
   a(t) &= \bigl(a^1(t),\, \ldots,\, a^N(t)\bigr)^\top, \\
   a^i(t) &= \sum_{j=1}^{n} \xi^{a_i}_j \, B_{j,k}(t),
\end{align}
where $N$ is the number of detectors and the $\xi^{a_i}_j$ are the learnable spline coefficients of detector $i$, which are standard-normally distributed. The basis functions $B_{j,k}$ of polynomial degree $k$ live on a fixed, uniformly spaced knot vector $\mathbf{t} = (t_0, \ldots, t_{n+k})$ and are defined recursively via the Cox--de Boor relation \citep{cox1972, deboor1972},
\begin{align} \label{eq:bspline_basis}
   B_{j,0}(t) &= \begin{cases} 1 & \text{if } t_j \leq t < t_{j+1}, \\ 0 & \text{otherwise,} \end{cases} \\
   B_{j,k}(t) &= \frac{t - t_j}{t_{j+k} - t_j}\,B_{j,k-1}(t) + \frac{t_{j+k+1} - t}{t_{j+k+1} - t_{j+1}}\,B_{j+1,k-1}(t). \nonumber
\end{align}
We use cubic B-splines ($k = 3$), which are piecewise polynomials of degree three that are joined at the knot locations with $C^2$ continuity, yielding smooth atmospheric priors. The number of basis functions $n$ controls the temporal resolution of the model and is chosen small enough to capture the dominant low-frequency atmospheric fluctuations without introducing degeneracies with the map signal. Since each detector is modelled separately, this component also absorbs the slow instrumental drifts discussed in \cref{sec:measurement-eq}, which vary on similar timescales.

A large part of the atmospheric signal is driven by the changing line of sight: as the telescope scans, the boresight elevation $\theta(t)$ varies and the atmospheric emission grows with the airmass $\propto 1/\sin\theta(t)$, inducing a smooth, elevation-dependent trend that is common to all detectors. To capture this explicitly, we weight the B-spline basis by the normalized elevation
\begin{equation} \label{eq:rel_el}
   \tilde{\theta}(t) = \frac{\theta(t) - \theta_{\min}}{\theta_{\max} - \theta_{\min}} \in [0, 1],
\end{equation}
where $\theta_{\min}$ and $\theta_{\max}$ are the extremal boresight elevations over the scan, and duplicate the basis into an elevation-weighted pair,
\begin{equation} \label{eq:bspline_el}
   a^i(t) = \sum_{j=1}^{n} \Bigl[ \xi^{a_i}_{j,1}\, \tilde{\theta}(t) + \xi^{a_i}_{j,2}\, \bigl(1 - \tilde{\theta}(t)\bigr) \Bigr] B_{j,k}(t),
\end{equation}
with $2n$ standard-normally distributed coefficients per detector. The two sets of coefficients describe the atmospheric curve at the highest and at the lowest elevation of the scan, respectively, between which the model interpolates linearly, capturing the elevation gradient at little extra cost. As the gradient is tied to the boresight elevation rather than to a sky position, it further helps to disentangle the atmosphere from the static astronomical signal.

\begin{figure*}
\centering
   \includegraphics[width=17cm]{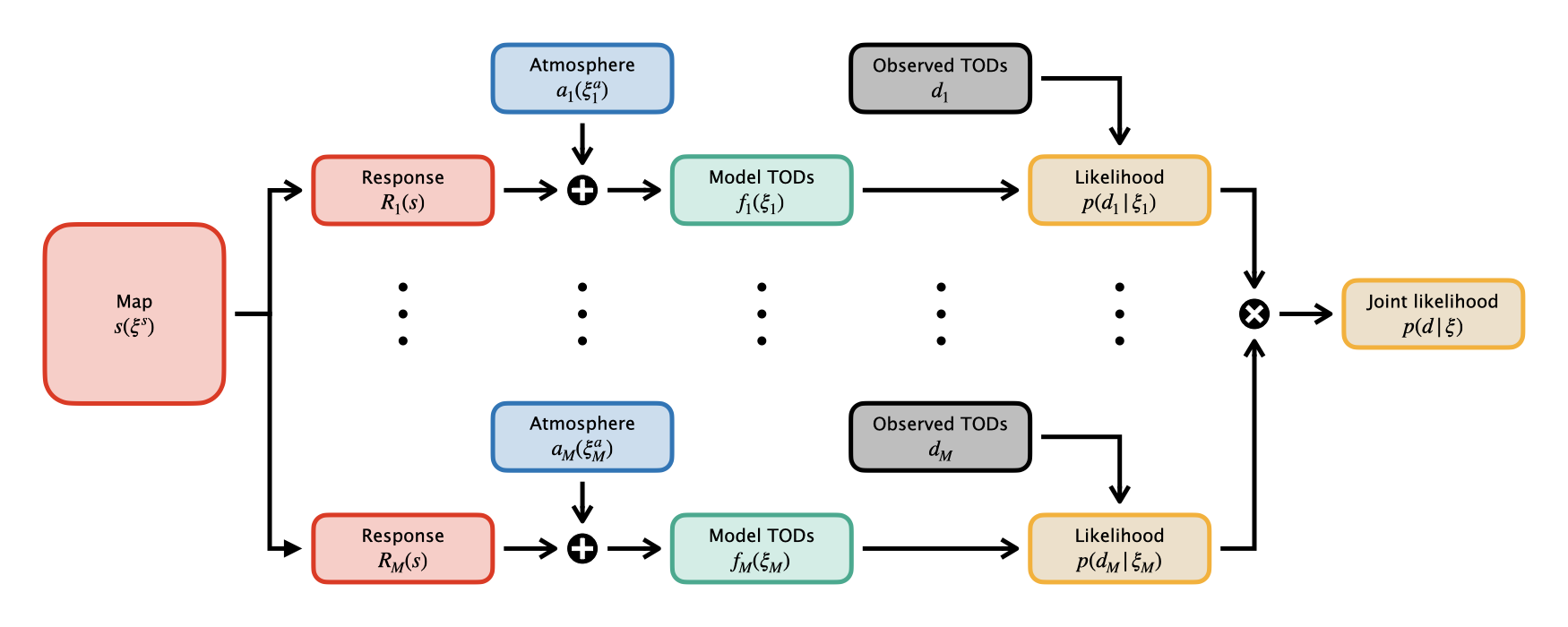}
     \caption{Visualization of the joint likelihood model used to infer a single map from multiple time-ordered datasets. Each likelihood term $p(d_m|\xi_m)$ for $m = 1, \ldots, M$ corresponds to different observed TODs $d_m$ with a separate scanning pattern $R_m$ applied to the map image. The individual model TODs $f_m(\xi_m)$ entering the likelihoods are described by different atmospheric contributions $a_m(\xi^a_m)$, while the underlying map model $s(\xi^s)$ is shared across all terms. The individual likelihoods are multiplied together to obtain the joint likelihood $p(d|\xi)$, which is the input to the variational inference method to infer the posterior distribution of the generative model parameters.}
     \label{fig:procedure}
\end{figure*}

\subsection{Likelihood} \label{sec:likelihood}

Having the prior models for the sky map and atmospheric emission at hand, we now derive the likelihood function for the observed data given the generative model parameters. Assuming that the instrumental noise $n$ is drawn from a Gaussian $\mathcal{G}(n,N)$ with diagonal covariance $N$ (uncorrelated noise between detectors), the likelihood is given by
\begin{equation} \label{eq:likelihood}
   p(d|\xi) = \mathcal{G}(d - f(\xi), N),
\end{equation}
where $f(\xi)$ is the generative model defined in \cref{eq:gen_model}. For numerical stability and computational efficiency, we work with the negative log-likelihood
\begin{equation} \label{eq:hamiltonian}
   H(d|\xi) = \frac{1}{2} (d - f(\xi))^\dagger N^{-1} (d - f(\xi)) + \text{const},
\end{equation}
also known as the information Hamiltonian (see, for example, \citealp{ensslin2019}). The constant term can be ignored during inference since it does not depend on the model parameters $\xi$.

For MUSTANG-2, the observations are usually split into multiple scans of the same region of the sky, each resulting in separate TODs with a distinct number of active detectors. For this purpose, we introduce a separate generative model $f_m(\xi_m)$ for each scan, which allows us to account for the different scanning patterns $R_m$ and atmospheric conditions $a_m(\xi^a_m)$ during each observation. We still assume the astronomical signal to be static across all scans by using a shared map model $s(\xi^s)$ for all datasets. Assuming that the individual time-ordered datasets $d_m$ and atmospheric contributions $a_m$ are independent, the joint likelihood is given by the product of the individual likelihoods, and the joint information Hamiltonian is given by the sum of the individual Hamiltonians,
\begin{align} \label{eq:joint-likelihood}
   p(d|\xi) &= \prod_m p(d_m|\xi_m), \\
   H(d|\xi) &= \sum_m H(d_m|\xi_m),
\end{align}
where each $p(d_m|\xi_m)$ and $H(d_m|\xi_m)$ is defined as in \cref{eq:likelihood} and \cref{eq:hamiltonian}, respectively. \Cref{fig:procedure} shows a visualization of this joint likelihood model with a shared map model. The likelihood serves as an input to the variational inference method used to estimate the posterior distribution of the generative model parameters. Equivalently, the scans can be combined into one large observation by concatenating the individual datasets $d_m$ and stacking the response functions $R_m$, which leads to exactly the same joint Hamiltonian, as shown in \cref{app:concatenation}.

\subsection{Posterior inference} \label{sec:inference}

To obtain the posterior distribution of the generative model parameters given the observed data, we need to compute the evidence $P(d)$ in \cref{eq:bayes}. For our problem, the posterior distribution is intractable due to the high-dimensional parameter space and the complexity of the generative model. Therefore, we use variational inference to approximate the posterior distribution $p(\xi|d)$ with a tractable distribution $q(\xi|d)$ by minimizing the Kullback--Leibler divergence 
\begin{equation}
   \mathcal{D}_{\text{KL}}(q(\xi|d) \| p(\xi|d)) = \int q(\xi|d) \log \frac{q(\xi|d)}{p(\xi|d)} d\xi
\end{equation}
between the two distributions. This turns the inference problem into an optimization problem, which can be solved with gradient-based methods.

In this work, we employ geometric variational inference (geoVI; \citealp{frank2021}), which approximates the posterior with a multivariate Gaussian distribution while taking into account the underlying geometry of the parameter space defined by the generative model. This allows accurate approximation of the posterior distribution even in cases where the true posterior is non-Gaussian. Furthermore, geoVI approximates the posterior via a set of samples that are iteratively updated to minimize the KL divergence, allowing reliable uncertainty quantification of the reconstructed quantities via sample averages \citep{ensslin2019}. The geoVI method is implemented in the \texttt{NIFTy} package and is designed to efficiently handle large-scale inference problems, making it well suited for our application of reconstructing sky maps from single-dish telescope data.

\section{Applications} \label{sec:applications}

We now apply our method to a MUSTANG-2 observation of the galaxy cluster RXJ1347, a massive galaxy cluster at a redshift of $z = 0.451$ and at one time the most X-ray luminous cluster known \citep{schindler1995}.\footnote{It was later surpassed as the most X-ray luminous cluster known by the SZ-selected Phoenix Cluster \citep{mcdonald2012}.} It hosts a cool core and is undergoing a merger, which imprints substructures on its SZ signal, making it an interesting test case for the multi-component map model of \texttt{maria-nifty}. The cluster has already been imaged at 90 GHz with MUSTANG, the 64-detector predecessor of MUSTANG-2 on the GBT \citep{mason2010, korngut2011}, and it has been part of a MUSTANG-2 study of compact sources in galaxy clusters \citep{dicker2021}, which reports a flux density of $3.31 \pm 0.34$ mJy at 90 GHz for the compact source at its centre.

\begin{figure}[h]
  \resizebox{\hsize}{!}{\includegraphics{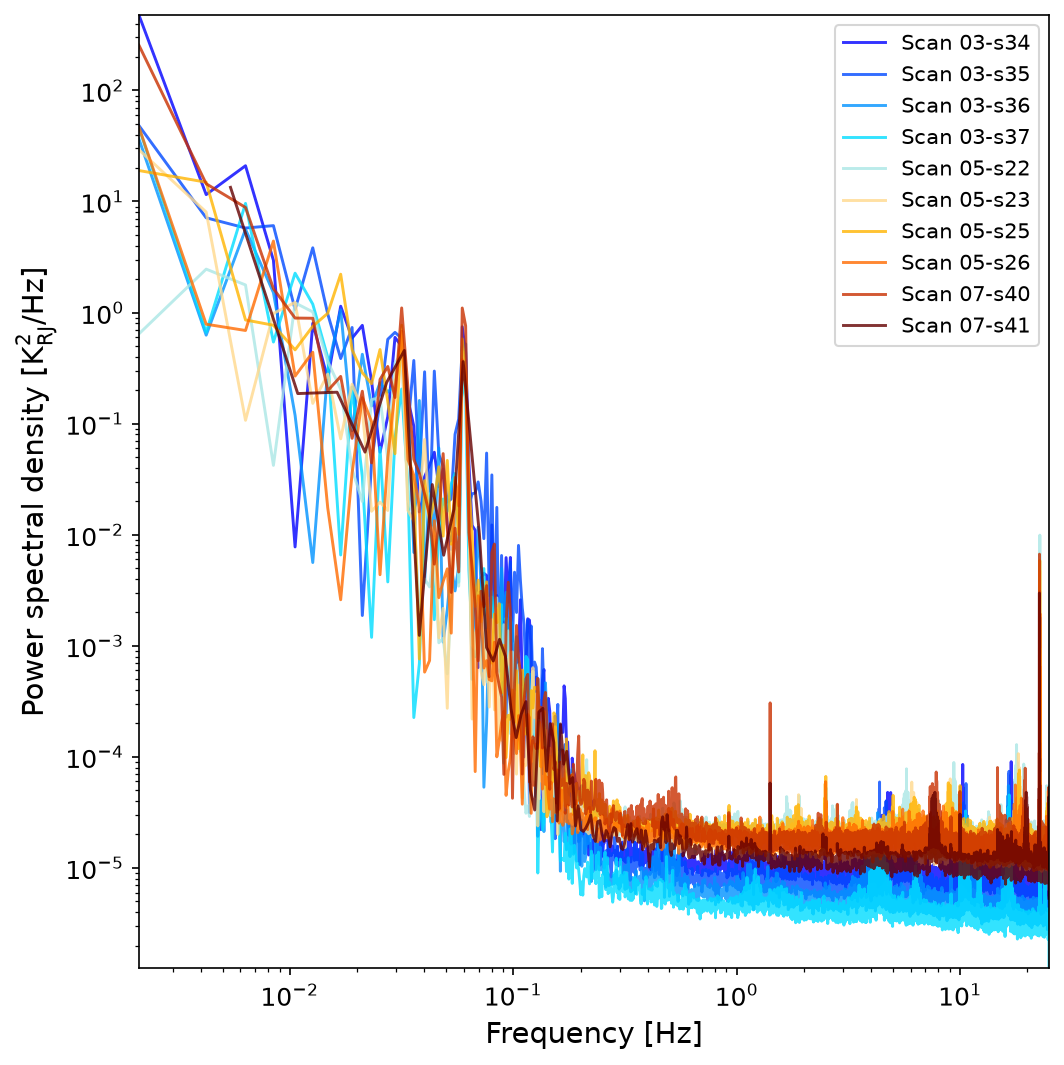}}
  \caption{Power spectra of the observed TODs for the ten scans of RXJ1347, coloured according to the scan. Below $\sim 0.1$ Hz, the spectra are dominated by the atmospheric emission, which falls off steeply towards higher frequencies. Above $\sim 0.2$ Hz, they flatten into the white noise floor of the instrument, on top of which several narrow lines are visible. The spectra of all scans agree closely, indicating comparable observing conditions across the three observing sessions.}
  \label{fig:data-spectra}
\end{figure}

Here, we use the publicly available MUSTANG-2 data of the GBT project AGBT17A\_358\footnote{\url{https://github.com/sievers/M2-TODs}}. They consist of ten TODs recorded in three observing sessions. Each observational scan comprises up to eight minutes of observing time using 142 to 166 active detectors. All scans are centred on the same position at a right ascension $\alpha = 13^{\rm h}47^{\rm m}30.5^{\rm s}$ and a declination $\delta = -11\degr45\arcmin09.0\arcsec$. \Cref{fig:data-spectra} shows the power spectra of the individual scans, which all exhibit a similar shape. At low frequencies, the atmosphere dominates and its power falls off steeply by several orders of magnitude, before the spectra flatten into the white noise floor of the instrument at high frequencies. The atmospheric conditions and noise levels are therefore comparable across the scans, allowing us to describe all of them with the same prior model and hyperparameters in a joint reconstruction. Further details on the individual scans are summarized in \cref{tab:rxj-tods}.

\begin{figure*} [t]
\centering
   \includegraphics[width=17cm]{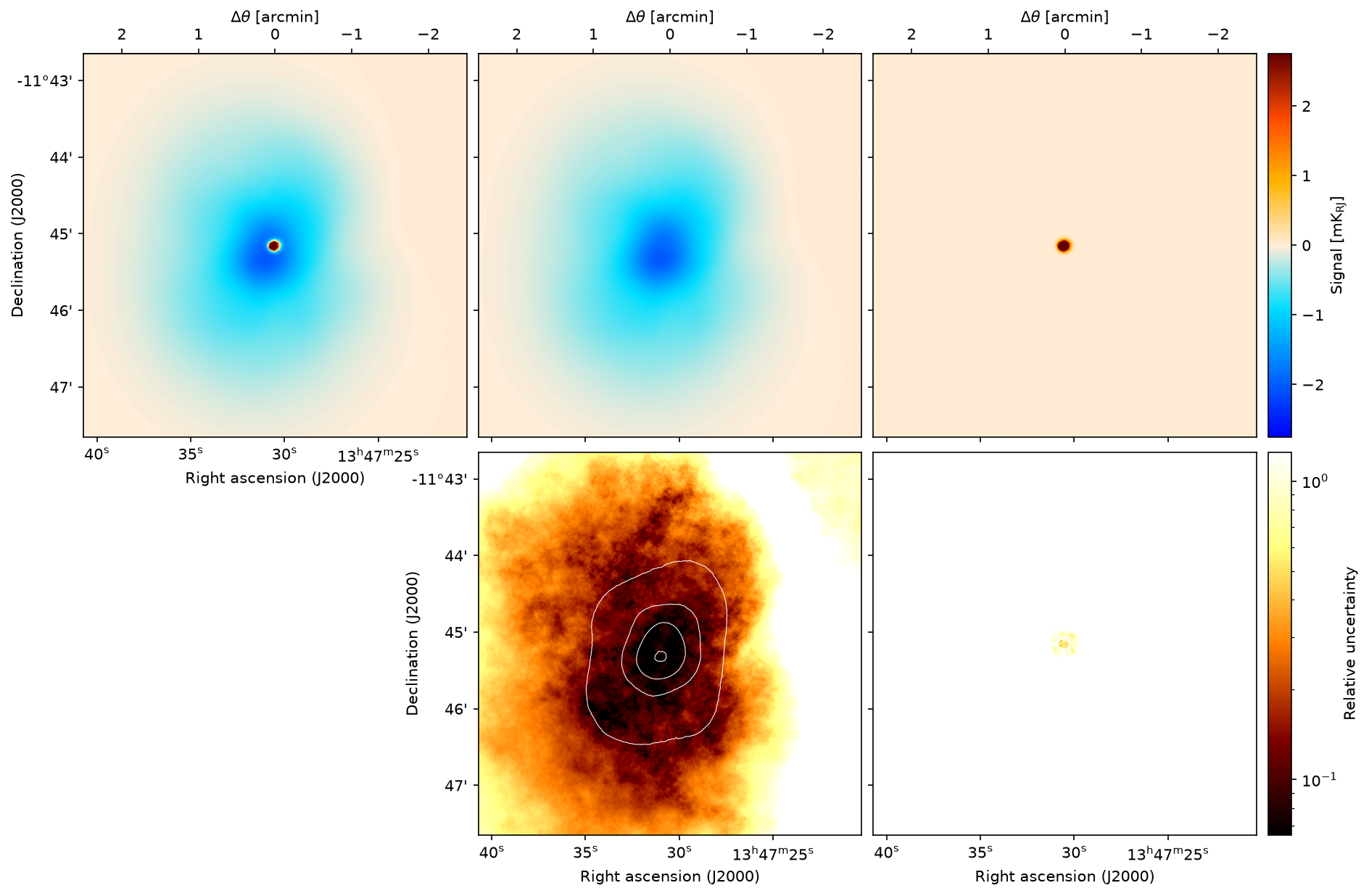}
     \caption{Map reconstruction of the MUSTANG-2 observation of RXJ1347. \textit{Top row, from left to right:} the posterior mean of the \texttt{maria-nifty} map model, followed by the posterior means of its two components, the extended SZ signal and the compact source, which add up to the full map. \textit{Bottom row:} the relative posterior uncertainty of the two components on a logarithmic scale, with the white contours indicating the reconstructed decrement of the panel above. It is smallest where the respective component carries signal and approaches unity where it vanishes. The temperature values of the compact source are clipped to keep a common colour scale for all panels of the top row. All panels show a cutout of the inner $5\arcmin \times 5\arcmin$ of the $10\arcmin \times 10\arcmin$ FOV of the map model, whose full extent is shown in \cref{fig:map-comparison}.}
     \label{fig:map-reconstruction}
\end{figure*}

\subsection{Model setup and inference} \label{sec:rxj-setup}

Since the observation consists of only ten datasets, we are able to inspect them individually and manually flag the jumps and large spikes they contain. No further preprocessing is applied.

We then set up the prior model introduced in \cref{sec:prior-model}, which consists of a map and an atmospheric component. For the atmosphere, we use the cubic spline model of \cref{eq:bspline_el} that additionally takes the boresight elevation of each scan into account. \Cref{fig:tod-atmosphere-prior} shows prior samples of the atmospheric model for each detector of one of the datasets.

Since RXJ1347 features a compact source next to its cool core, we describe the sky with a map model consisting of two different components. The first one is located at the scan centre and spans a FOV of $10\arcmin \times 10\arcmin$, which covers the extended SZ signal of the cluster. The second one is also centred at the scan centre and covers a much smaller FOV to capture the compact source. While the extended component is free to take both positive and negative values, the compact source is restricted to positive values. Their FOVs, resolutions, and centre shifts are listed in \cref{tab:rxj-components}, while the CFM hyperparameters can be found in \cref{tab:params}. The hyperparameters are chosen such that their prior amplitudes are comparable in magnitude. \Cref{fig:map_prior} visualizes sixteen prior samples of this two-component map model, showing the flexibility of the employed map prior. 

We emphasize that the widths of the Gaussians of the map priors are learnable, such that the extent of both components is inferred from the data instead of being fixed a priori. For the compact source, the two-dimensional envelope alone could already describe a Gaussian source, while the CFM additionally allows for any deviation from a pure Gaussian structure. For the extended SZ signal, in turn, the window concentrates the signal towards the centre of the FOV, but the model can still capture emission beyond the extent of the prior due to the learnable width of the Gaussian.

With the prior models in place, we set up the joint likelihood of \cref{eq:joint-likelihood} for the ten datasets and infer all model parameters simultaneously. For the inference, we run the geoVI algorithm with eight posterior samples.

\begin{table}[h]
\caption{Map prior components.}
\label{tab:rxj-components}
\centering
\begin{tabular}{l c c c}
\hline\hline
\noalign{\vskip 4pt}
Component & FOV & Resolution & Centre shift \\
\noalign{\vskip 4pt}
\hline
\noalign{\vskip 4pt}
   SZ signal & $10\arcmin \times 10\arcmin$ & $1000 \times 1000$ & $(0, 0)$ \\
\noalign{\vskip 4pt}
   Compact source & $0.5\arcmin \times 0.5\arcmin$ & $50 \times 50$ & $(0, 0)$ \\
\noalign{\vskip 2pt}
\hline
\end{tabular}
\end{table}

\begin{figure*} [t]
\centering
   \includegraphics[width=17cm]{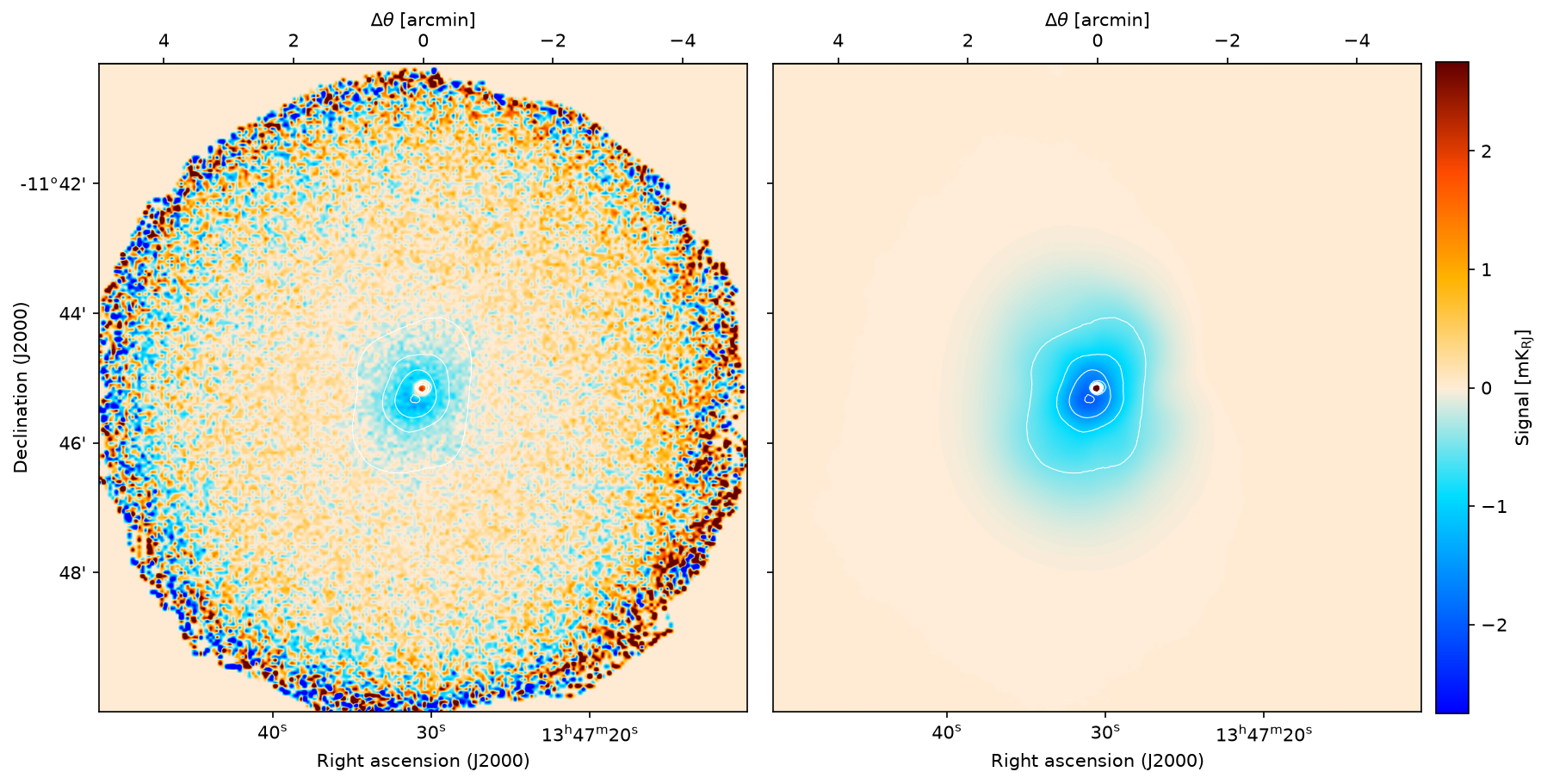}
     \caption{Comparison of two reconstructions of the same MUSTANG-2 observation of RXJ1347. \textit{Left:} a \texttt{minkasi} reconstruction of the data following the procedures in \cite{romero2020} and \cite{orlowskischerer2022}. \textit{Right:} the posterior mean of the \texttt{maria-nifty} map model. Both panels share the colour scale on the right. While both recover the thermal SZ decrement of the cluster and the compact source at its centre, the \texttt{minkasi} map is dominated by noise on small scales, whereas our reconstruction is considerably smoother due to the GP prior of the map. Both panels show the map at a FOV of $10\arcmin \times 10\arcmin$.}
     \label{fig:map-comparison}
\end{figure*}

\subsection{Imaging results} \label{sec:rxj-results}

The top row of \cref{fig:map-reconstruction} shows the map reconstruction obtained after the inference with geoVI, together with the reconstructions of the two individual map components. All of them depict the posterior mean of the map, which is calculated from the individual posterior samples in \cref{fig:map-posterior}. Our method achieves a clean separation of the individual map components, even though they are superposed on the sky. The SZ signal of the cluster manifests as a decrement (negative signal relative to the CMB) around 93~GHz, which is captured by the negative surface brightness in the extended component. On top of it, the positive contribution of the compact source is visible, which is recovered by the second component. It reaches peak values of up to $30\,\mathrm{mK}_{\mathrm{RJ}}$ and an integrated flux density of $3.11 \pm 0.33\,\mathrm{mJy}$, in agreement with the $3.31 \pm 0.34\,\mathrm{mJy}$ reported by \citet{dicker2021}. The morphology of the reconstruction is moreover consistent with ALMA observations of the cluster core \citep{kitayama2016}, which show the deepest part of the decrement south-east of the central AGN, as discussed in \cref{app:alma}.

The bottom row of \cref{fig:map-reconstruction} shows the relative posterior uncertainties of the two map components. For the extended SZ signal, it is lowest in the centre of the cluster, where the signal is strongest, and increases towards the edges of the map, where the scanning pattern provides less data compared to the map centre. The relative uncertainty of the SZ component reaches values of $\sim 10\%$ at the cluster centre and remains comparatively low across the central $\sim 3\arcmin$, which corresponds to the well-sampled region of the daisy scan. Further out, it gradually increases towards unity, so that the reconstruction there is increasingly informed by the prior. Part of this increase is due to the reconstructed signal approaching zero, which inflates the relative uncertainty even where the absolute uncertainty remains small. The uncertainty of the compact source is higher compared to the decrement, since the convolution with the instrument beam makes it difficult to constrain the exact peak value and shape of the source.

\subsection{Comparison to a \texttt{minkasi} reconstruction}

As the map-maker most commonly used in recent MUSTANG-2 studies, \texttt{minkasi} serves as the baseline reconstruction method here. It follows the maximum-likelihood approach and treats the atmospheric emission as correlated noise. Given the response $R$ and a time-domain noise covariance $N$ that accounts for both the atmosphere and the instrumental noise, the maximum-likelihood map
\begin{equation} \label{eq:ml_map}
   \hat{s} = \left(R^\dagger N^{-1} R\right)^{-1} R^\dagger N^{-1} d
\end{equation}
is obtained iteratively with a preconditioned conjugate gradient algorithm, while the noise covariance is estimated from the data themselves \citep{romero2020}. To this end, the TODs of each scan are rotated into the eigenbasis of the detector--detector covariance obtained from a singular value decomposition, and the power spectrum of each eigenmode is estimated and smoothed. The resulting $N$ is thus stationary in time but not diagonal across detectors, so that it captures the correlations between detectors introduced by the atmosphere, which are dominated by the leading modes. The \texttt{minkasi} map used here was reconstructed from the same ten scans, using 20 conjugate gradient iterations. For the comparison, we cut out the same $10\arcmin \times 10\arcmin$ region as our extended map component, which corresponds to a resolution of $200 \times 200$ pixels.

\Cref{fig:map-comparison} shows a comparison of the maps reconstructed using \texttt{maria-nifty} versus that from \texttt{minkasi} following the established MUSTANG-2 map-making procedures from e.g.\ \cite{romero2020} and \cite{orlowskischerer2022}.
We note that our application of \texttt{minkasi} here is simply to provide a baseline for comparison and that \texttt{minkasi} provides a flexible framework that could likely include priors similar to those implemented in \texttt{maria-nifty}. Extensions to \texttt{minkasi} will be presented in future work.

Both the \texttt{maria-nifty} and \texttt{minkasi} maps recover the thermal SZ decrement of the cluster and the compact source at its centre. Since the CFM prior correlates neighbouring pixels, we obtain a considerably smoother map that features fewer artefacts than the \texttt{minkasi} reconstruction, which is dominated by noise on small scales. 
The noise level of the \texttt{minkasi} map moreover increases strongly towards the outer parts of the map, forming a ring of enhanced fluctuations around the well-covered centre. There, the daisy pattern provides almost no data, and a maximum-likelihood estimate of these pixels is essentially unconstrained. In our reconstruction, the same lack of data is instead absorbed by the prior.

\begin{figure}[t]
  \resizebox{\hsize}{!}{\includegraphics{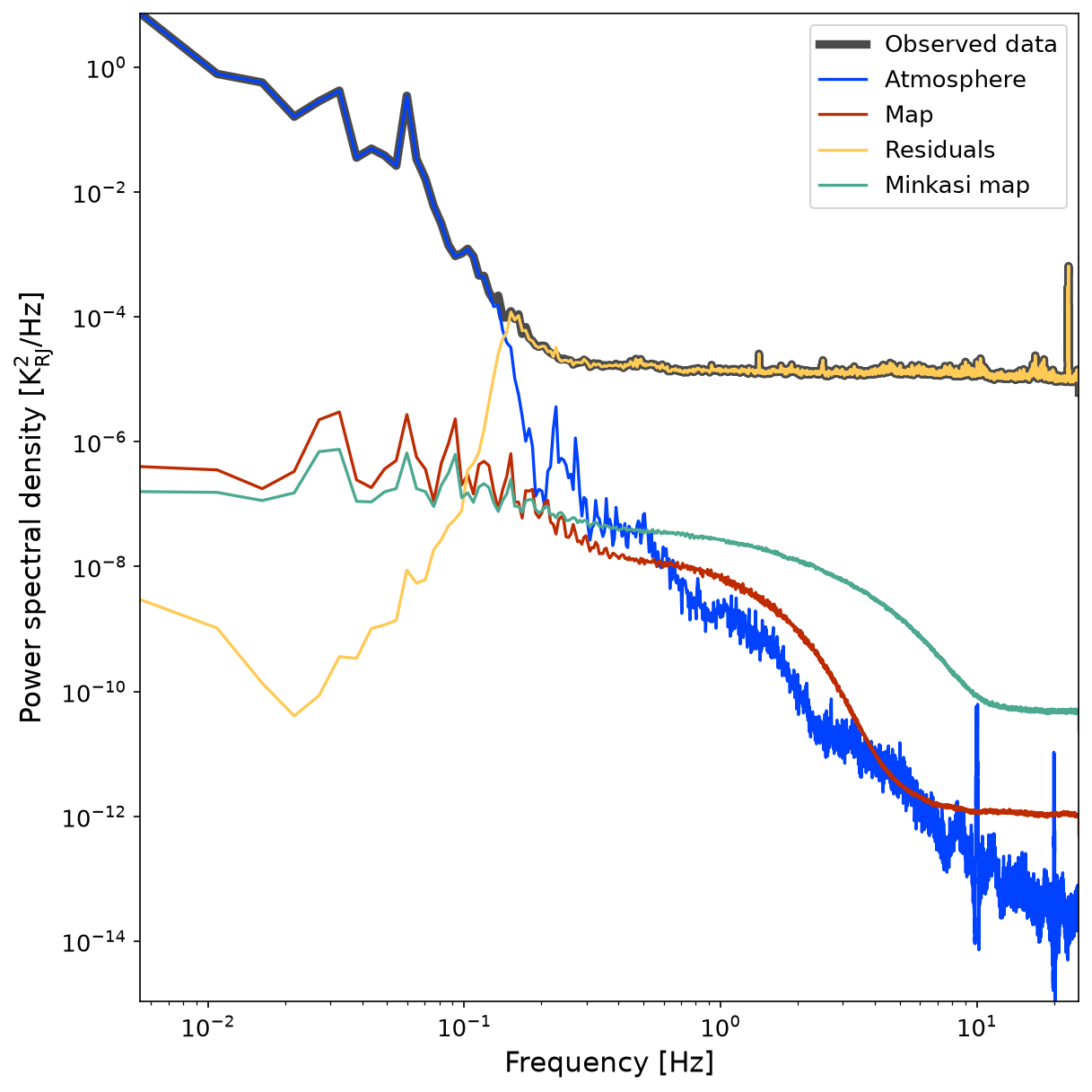}}
  \caption{Average power spectra of the observed data and of the individual contributions of the forward model shown in \cref{fig:tod-components}. Below $\sim 0.1$ Hz, the spectrum of the observed data is entirely dominated by the atmospheric emission, while above $\sim 0.2$ Hz it is dominated by the white noise, which is recovered by the residuals. Around the transition between the two regimes ($\sim 0.15$ Hz), the residuals rise slightly above the white noise level, indicating low-frequency drifts that are not fully captured by the spline model of the atmosphere. The map response contributes the least power on all timescales, staying about an order of magnitude below the noise level. For comparison, the \texttt{minkasi} map is projected into the time domain with the same response function.}
  \label{fig:component-spectra}
\end{figure}

The same behaviour is visible in the power spectra of \cref{fig:component-spectra}. Below $\sim 0.2$ Hz, our map model carries more power than the \texttt{minkasi} map, which corresponds to the large angular scales of the extended SZ signal that are partly removed by the filtering of the traditional map-making. This is also reflected in the maps themselves, where we recover a deeper SZ decrement than the \texttt{minkasi} reconstruction, as shown by the difference of the two maps in \cref{fig:map-difference}. The compact source reaches a considerably higher peak value in our reconstruction, while its integrated flux is comparable. After convolution with the beam, the difference at the position of the source largely vanishes, so the higher peak is mainly a consequence of the finer resolution of our map model rather than of additional flux. Above $\sim 0.5$ Hz, the ordering reverses and the \texttt{minkasi} map exceeds our map model by more than an order of magnitude, corresponding to the small angular scales sampled during the scan. This excess is consistent with the small-scale noise that dominates the \texttt{minkasi} map, while the CFM prior suppresses such fluctuations by correlating neighbouring pixels.  

Moreover, \texttt{minkasi} follows the maximum-likelihood approach and returns only a single best-fit map, whereas our approach approximates the full posterior distribution via a set of samples, from which we derive the posterior mean and the relative standard deviation, which quantifies the uncertainties of our reconstructed maps.

A further difference lies in how much of the observation is modelled explicitly. While \texttt{minkasi} reconstructs a single map that contains all astronomical emission, our map model consists of two components, which separate the extended SZ signal from the compact source at the cluster centre, although both overlap on the sky. The same holds for the atmosphere, which is absorbed into the noise covariance of \texttt{minkasi}, whereas our model describes it as a separate component with its own prior that is reconstructed alongside the map. As each detector is described by its own spline realization, this component can follow the non-stationary behaviour of the perceived noise discussed in \cref{sec:measurement-eq} to some degree, which a single, stationary noise covariance cannot capture.

\subsection{Contributions of the model components}

The separate modelling of the atmosphere and sky map components allows us to inspect the individual contributions of our forward model and to see how they add up to the observed data. \Cref{fig:component-spectra} shows the average power spectra of the individual model components and the residuals between model and data. At low frequencies (below $\sim 0.1$ Hz), the observed data are dominated by the atmosphere, whose power exceeds that of the map by several orders of magnitude, while white noise dominates at high frequencies (above $\sim 0.2$ Hz), which constitutes a substantial fraction of the residuals. The map stays about an order of magnitude below this noise level and is thus subdominant across the entire spectrum. It can nevertheless be recovered, since it is tied to a fixed position on the sky and is sampled repeatedly by the different detectors and scans.

\begin{figure*}[t]
\centering
   \includegraphics[width=17cm]{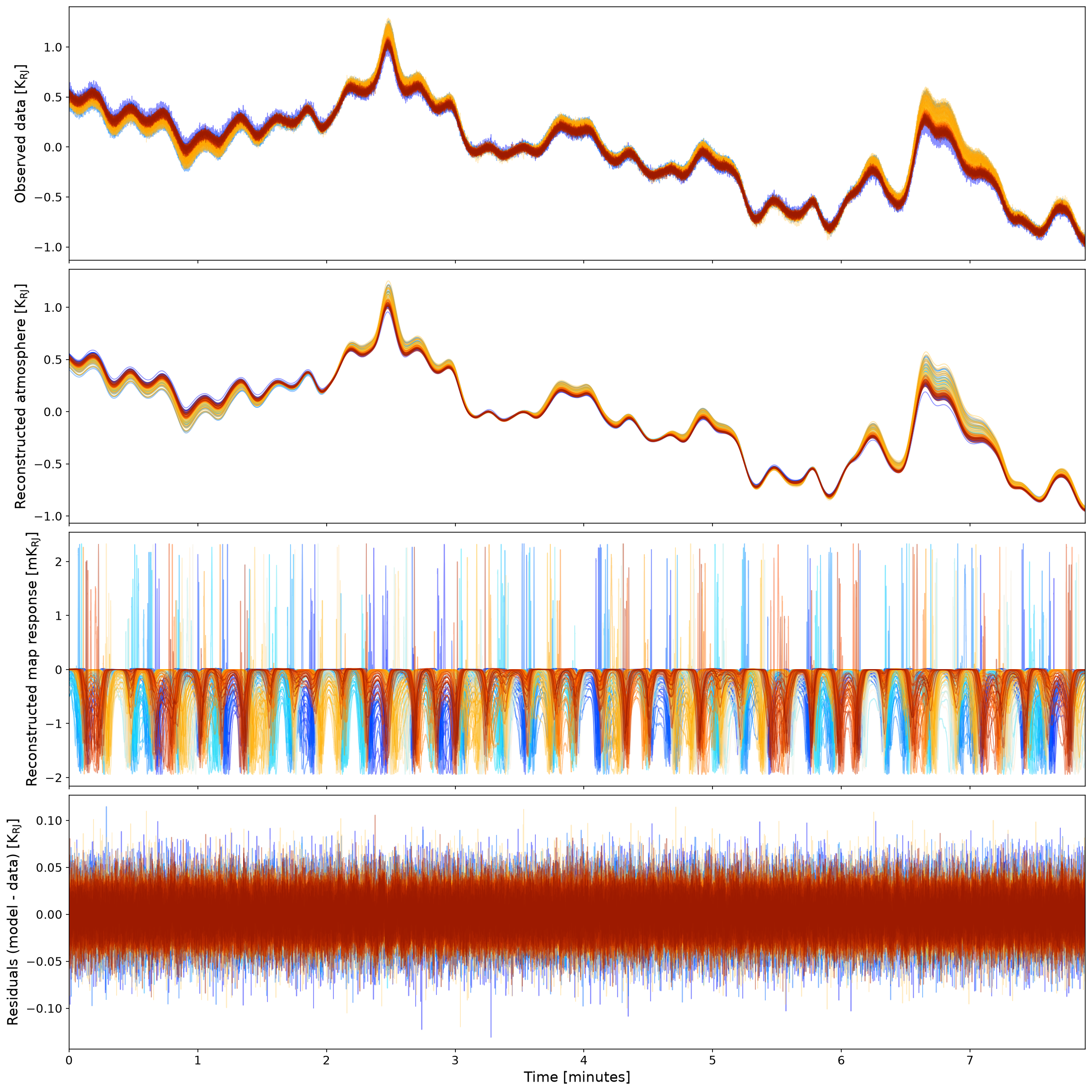}
     \caption{Contributions of the forward model in data space for a single scan of RXJ1347, with the detectors coloured according to their index. \textit{From top to bottom:} the observed TODs, the reconstructed atmospheric emission, the reconstructed map response, and the residuals between model and data. Note the different units of the individual panels. While the observed data and the atmosphere are of the order of one Kelvin, the map response only reaches a few mK, showing negative dips whenever a detector crosses the SZ signal of the cluster and positive spikes at the position of the compact source. The residuals show little remaining structure, indicating that the model describes the data down to the instrumental noise.}
     \label{fig:tod-components}
\end{figure*}

The same hierarchy is visible in the time domain, shown in \cref{fig:tod-components} for one of the ten scans. The observed TODs are almost entirely explained by the reconstructed atmosphere, which follows their large-scale fluctuations of the order of one Kelvin. The reconstructed map response is three orders of magnitude smaller, with negative dips whenever a detector crosses the SZ signal of the cluster and positive spikes at the position of the compact source. The residuals show little structure and are consistent with the instrumental noise, indicating that our model describes the data down to the noise level.

A closer inspection of the power spectra in \cref{fig:component-spectra}, however, reveals that the instrumental noise is not perfectly white. Around the transition between the two regimes (at $\sim 0.15$ Hz), the residuals rise above the white noise level. Part of the low-frequency drifts is therefore not captured by the spline model, so that the assumption of a time-independent noise in \cref{eq:mse} only holds approximately. Describing the remaining correlated noise explicitly, for example with an additional time-dependent noise component or an additional atmospheric contribution, is left for future work.

\section{Conclusions} \label{sec:conclusions}

In this work, we applied Gaussian process-based imaging to real (sub-)mm single-dish telescope data. We extended the multi-component approach described in \citet{wuerzinger2025}, which describes the sky map and the atmospheric emission as separate components of one generative model, and adapted it to the properties of MUSTANG-2 observations. To reconstruct a single map from several scans of the same field, we combined their likelihoods into a joint likelihood with a shared map model and inferred all model parameters simultaneously using variational inference.

To demonstrate our method, we applied it to the publicly available MUSTANG-2 observation of the galaxy cluster RXJ1347, which consists of ten scans. Since the cluster hosts a compact source next to its cool core, we described the sky with a two-component map model, which cleanly separated the extended thermal SZ decrement from the compact source. We compared our result to a \texttt{minkasi} reconstruction of the same data, which follows the maximum-likelihood approach and treats the atmospheric emission as correlated noise. Since the correlated field prior correlates neighbouring pixels, we obtained a considerably smoother map with fewer artefacts, together with the per-pixel uncertainty of the recovered signal. Moreover, the atmosphere is reconstructed alongside the map, allowing us to inspect the individual contributions of the forward model, which describe the observed data down to the noise level.

Several extensions of the method appear promising. A natural next step is to scale these methods to the large datasets expected for AtLAST, whose two-degree field of view is designed to illuminate up to $\mathcal{O}(10^7)$ detectors per frequency band \citep{gallardo2024, mroczkowski2025}, i.e., many orders of magnitude more than the 223 detectors of MUSTANG-2. The method can already be tested on realistic simulations of these arrays with \texttt{maria}, which additionally allows the reconstructions to be validated against a known ground truth. Our joint likelihood is already built for such arrays, since the concatenation of multiple datasets and stacking of the response functions and atmospheric contributions is equivalent to having one large array with a high number of detectors (see \cref{app:concatenation}). However, the main limitation for larger observations is the computational cost of the inference, especially for multiple scans. As \texttt{NIFTy} is written in \texttt{JAX}, the inference could be distributed across several GPUs, either by splitting the datasets or the posterior samples over the available devices. This would allow reconstructions from the much larger detector numbers that these facilities will deliver.

Another possible extension is multi-frequency imaging, which would allow us to separate the astronomical components not only by their spatial structure, but also by their spectral behaviour. Following the spatio-spectral models of \texttt{NIFTy} used in other fields such as X-ray \citep{westerkamp2024, eberle2024, guardiani2025} and $\gamma$-ray imaging \citep{scheel2023}, the map components could be equipped with an explicit spectral model that is inferred jointly with the sky brightness. Since the atmospheric emission follows its own frequency dependence, the additional bands would also constrain the separation of map and atmosphere.

Finally, the sky model and the instrument response are fully decoupled in our framework, so that other instruments can be included by exchanging the response function. Together with a spectral model, several instruments could contribute to a single reconstruction, each constraining its own range of angular scales and frequencies. Examples are single-dish telescopes with different dish sizes, such as the 6-meter Atacama Cosmology Telescope (ACT; \citealp{dunner2013}) and the 100-meter GBT, or the combination of a single-dish telescope with a radio interferometer. Since \texttt{NIFTy} is already used for radio interferometric imaging \citep{arras2022, roth2024, fuchs2026}, their responses could simply be added to our joint likelihood. In this way, \texttt{maria-nifty} offers a natural starting point for combining the data of current and upcoming facilities into a single, consistent reconstruction of the (sub-)millimetre sky.

\begin{acknowledgements}

The GBT data were acquired under the project ID AGBT17A\_358 (PI Brian Mason).  The Green Bank Observatory is a facility of the National Science Foundation operated under cooperative agreement by Associated Universities, Inc. 

R.F. acknowledges funding through the German Federal Ministry of Education and Research for the project ErUM-IFT: Informationsfeldtheorie für Experimente an Großforschungsanlagen (Förderkennzeichen: 05D23EO1).

L.H. and J.W. are supported by the Excellence Cluster ORIGINS, which is funded by the Deutsche Forschungsgemeinschaft (DFG, German Research Foundation) under Germany’s Excellence Strategy -- EXC-2094-390783311.

T.M.\ acknowledges support from the Agencia Estatal de Investigaci\'on (AEI) and the Ministerio de Ciencia, Innovaci\'on y Universidades (MICIU) Grant ATRAE2024-154740 funded by MICIU/AEI//10.13039/501100011033.
T.M.\ is also partly supported by the Spanish program Unidad de Excelencia María de Maeztu CEX2020-001058-M, financed by MCIN/AEI//10.13039/501100011033, and by the MaX-CSIC Excellence Award MaX4-SOMMA-ICE. 

S.D.\ and the MUSTANG-2 instrument were supported by the National Science Foundation under Grant Nos.\ 1509093 and 1309032.

AtLAST has received funding from the European Union's Horizon Europe and Horizon2020 research and innovation programmes under grant agreements No.\ 101188037 and No.\ 951815.
Views and opinions expressed are however those of the authors only and do not necessarily reflect those of the European Union or European Research Executive Agency. Neither the European Union nor the European Research Executive Agency can be held responsible for them.

\end{acknowledgements}

\bibliographystyle{aa}
\bibliography{references}

\begin{appendix}
\crefalias{section}{appendix}

\section{Equivalence of the joint likelihood and data concatenation} \label{app:concatenation}

Instead of multiplying the individual likelihoods as in \cref{eq:joint-likelihood}, the $M$ scans can be treated as a single, large observation by concatenating the datasets and stacking the response functions and atmospheric contributions,
\begin{align} \label{eq:concatenation}
   d &= (d_1^\dagger, \ldots, d_M^\dagger)^\dagger, &
   R &= (R_1^\dagger, \ldots, R_M^\dagger)^\dagger, \\
   a &= (a_1^\dagger, \ldots, a_M^\dagger)^\dagger, &
   N &= \text{diag}(N_1, \ldots, N_M),
\end{align}
such that the joint generative model reads
\begin{equation} \label{eq:joint-gen-model}
   f(\xi) = R \, s(\xi^s) + a(\xi^a)
   ~~\text{with}~~ \xi = \{\xi^s, \xi^{a_1}, \ldots, \xi^{a_M}\} .
\end{equation}
Stacking the individual response functions into a single operator is possible because the same map model $s(\xi^s)$ enters every scan. This is the case since we assume the astronomical signal to be static across all scans. The joint noise covariance $N$, in turn, is block diagonal since the individual datasets are assumed to be independent.

The residual of the concatenated measurement decomposes blockwise into the residuals of the individual scans,
\begin{equation} \label{eq:residual-blocks}
   r \equiv d - f(\xi) = (r_1^\dagger, \ldots, r_M^\dagger)^\dagger
   ~~\text{with}~~ r_m = d_m - f_m(\xi_m) ,
\end{equation}
where the $m$-th block of $R \, s(\xi^s)$ is given by $R_m \, s(\xi^s)$ and the $m$-th block of $a(\xi^a)$ by $a_m(\xi^{a_m})$. Furthermore, the inverse of a block-diagonal matrix is block diagonal with the individual blocks inverted,
\begin{equation}
   N^{-1} = \text{diag}(N_1^{-1}, \ldots, N_M^{-1}),
\end{equation}
such that all cross terms between different scans vanish in the quadratic form of \cref{eq:hamiltonian},
\begin{equation} \label{eq:no-cross-terms}
   r^\dagger N^{-1} r = \sum_{m,m'} r_m^\dagger \left(N^{-1}\right)_{mm'} r_{m'} = \sum_m r_m^\dagger N_m^{-1} r_m .
\end{equation}
The normalization term of the Gaussian splits in the same way, since the determinant of a block-diagonal matrix factorizes into $\log|2\pi N| = \sum_m \log|2\pi N_m|$. Together, this yields
\begin{align} \label{eq:hamiltonian-splits}
   H(d|\xi) &= \frac{1}{2} \sum_m \left( r_m^\dagger N_m^{-1} r_m + \log|2\pi N_m| \right) \nonumber \\
            &= \sum_m H(d_m|\xi_m) ,
\end{align}
which is exactly the joint information Hamiltonian of \cref{eq:joint-likelihood}. Therefore, the two Hamiltonians are identical as functions of $\xi$, and together with the same prior $p(\xi)$, both formulations define the same posterior distribution.

This equivalence holds for the model considered in this work, but it breaks down in a few cases. If the noise is correlated between the scans, the joint covariance $N$ is no longer block diagonal and the cross terms in \cref{eq:no-cross-terms} contribute, such that the Hamiltonian does not factorize. The same is true for the atmosphere, which we describe by a separate model for each scan. A single global model with correlations between consecutive scans would share its parameters across the individual likelihood terms. The static map, finally, is what allows us to stack the response functions in the first place. A source that varies on the timescale of the observations would instead require a time-variable map model, such that the input to the individual response functions changes from scan to scan and the stacking described above no longer applies.

As long as these assumptions hold, which of the two formulations to use is purely a matter of implementation. In \texttt{maria-nifty}, we tested both of them and obtained the same reconstruction results. Nevertheless, the concatenated formulation is preferable, since it harmonizes better with the vectorization functionality of \texttt{JAX} and therefore runs faster and more efficiently. This becomes particularly important for observations that consist of a large number of scans.

\section{RXJ1347 data} \label{app:rxj-data}

\Cref{tab:rxj-tods} summarizes the ten datasets of the GBT project AGBT17A\_358 that we use for the reconstruction of RXJ1347. They were recorded during three observing sessions, which are labelled by the session number and the scan number of the individual TODs. For each of them, we list the number of active detectors $N_{\mathrm{dets}}$, the number of time steps $N_{\mathrm{steps}}$, and the corresponding observing time. The number of active detectors varies between the scans, since detectors that are not working properly are excluded from the data. Furthermore, all scans cover about eight minutes of observing time, except for the last one, which is considerably shorter.

\begin{table}[h]
\caption{Overview of the RXJ1347 datasets.}
\label{tab:rxj-tods}
\centering
\begin{tabular}{lccc}
\hline\hline
\noalign{\vskip 4pt}
TODs & $N_{\mathrm{dets}}$ & $N_{\mathrm{steps}}$ & Time [min:ss] \\
\noalign{\vskip 4pt}
\hline
\noalign{\vskip 4pt}
   AGBT17A\_358\_03-s34 & 166 & 23691 & 7:54 \\
   AGBT17A\_358\_03-s35 & 160 & 23691 & 7:54 \\
   AGBT17A\_358\_03-s36 & 163 & 23690 & 7:54 \\
   AGBT17A\_358\_03-s37 & 163 & 23690 & 7:54 \\
   AGBT17A\_358\_05-s22 & 150 & 23690 & 7:54 \\
   AGBT17A\_358\_05-s23 & 151 & 23691 & 7:54 \\
   AGBT17A\_358\_05-s25 & 142 & 23690 & 7:54 \\
   AGBT17A\_358\_05-s26 & 145 & 23691 & 7:54 \\
   AGBT17A\_358\_07-s40 & 150 & 23690 & 7:54 \\
   AGBT17A\_358\_07-s41 & 153 &  9218 & 3:04 \\
\noalign{\vskip 4pt}
\hline
\end{tabular}
\end{table}

\section{Model hyperparameter priors} \label{app:params}

The map components are described by the CFM, whose hyperparameter priors are listed in \cref{tab:params} as pairs of mean and standard deviation. The mean of the sky brightness for the SZ signal is set to zero, as the map is expected to fluctuate around zero once the atmospheric contribution is separated from it. For the compact source, we enforce positivity via the softplus function and initialize the component with a low mean of $\log(1+e^{-10}) \approx 5 \times 10^{-5}$. Zero mode and fluctuations set the overall amplitude of the field and define how much it can deviate from the mean. The slope of the power spectrum of $-2$ favours smooth, spatially correlated structures. The last two entries (sigma x and sigma y) define the width of the two-dimensional Gaussian multiplying the CFM output, given in units of the FOV of the respective component. For the exact definitions of the CFM hyperparameters, see \citet{arras2022}.

\begin{table}[t]
\caption{Hyperparameter priors for the correlated field model.}
\label{tab:params}
\centering
\begin{tabular}{l cc cc}
\hline\hline
\noalign{\vskip 4pt}
Component & \multicolumn{2}{c}{SZ signal} & \multicolumn{2}{c}{Compact source} \\
& mean & std & mean & std \\
\noalign{\vskip 4pt}
\hline
\noalign{\vskip 4pt}
   mean & 0 & - & -10 & - \\
   zero mode & 0.001 & 0.01 & 1 & 0.5 \\
   fluctuations & 0.001 & 0.01 & 1 & 0.5 \\
   slope & -2 & 0.5 & -2 & 0.2 \\
   sigma x & 0.1 & 0.01 & 0.1 & 0.01 \\
   sigma y & 0.1 & 0.01 & 0.1 & 0.01 \\

\noalign{\vskip 4pt}
\hline
\end{tabular}
\end{table}

\section{Beam-convolved map comparison} \label{app:smoothed-map}

\begin{figure}[t]
  \resizebox{\hsize}{!}{\includegraphics{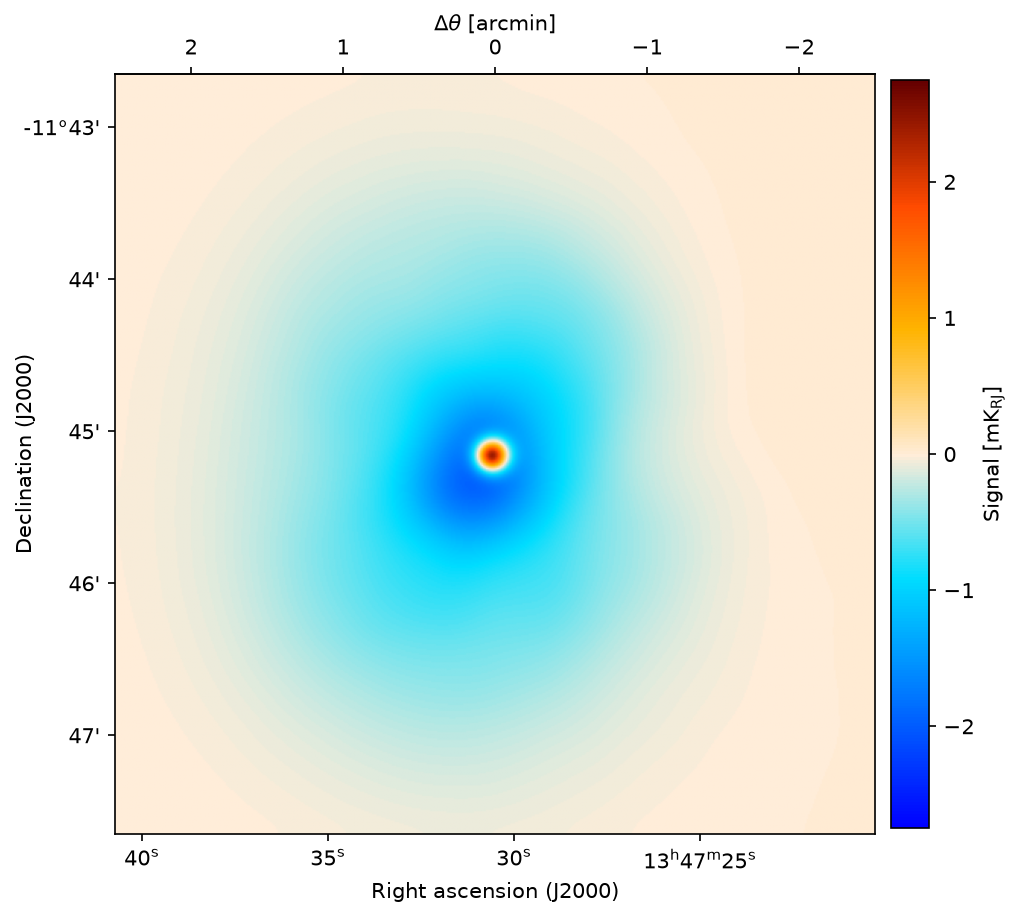}}
  \caption{Posterior mean of the full map model, convolved with the Gaussian beam of MUSTANG-2 and shown for the same cutout as \cref{fig:map-reconstruction}. The panel contains the sum of the two map components, with the SZ decrement of the cluster in blue and the compact source at its centre in red. The convolution shows the reconstruction at the angular resolution at which it is constrained by the data, in contrast to the finer resolution of the map model itself.}
  \label{fig:map-smoothed}
\end{figure}

\begin{figure}[t]
  \resizebox{\hsize}{!}{\includegraphics{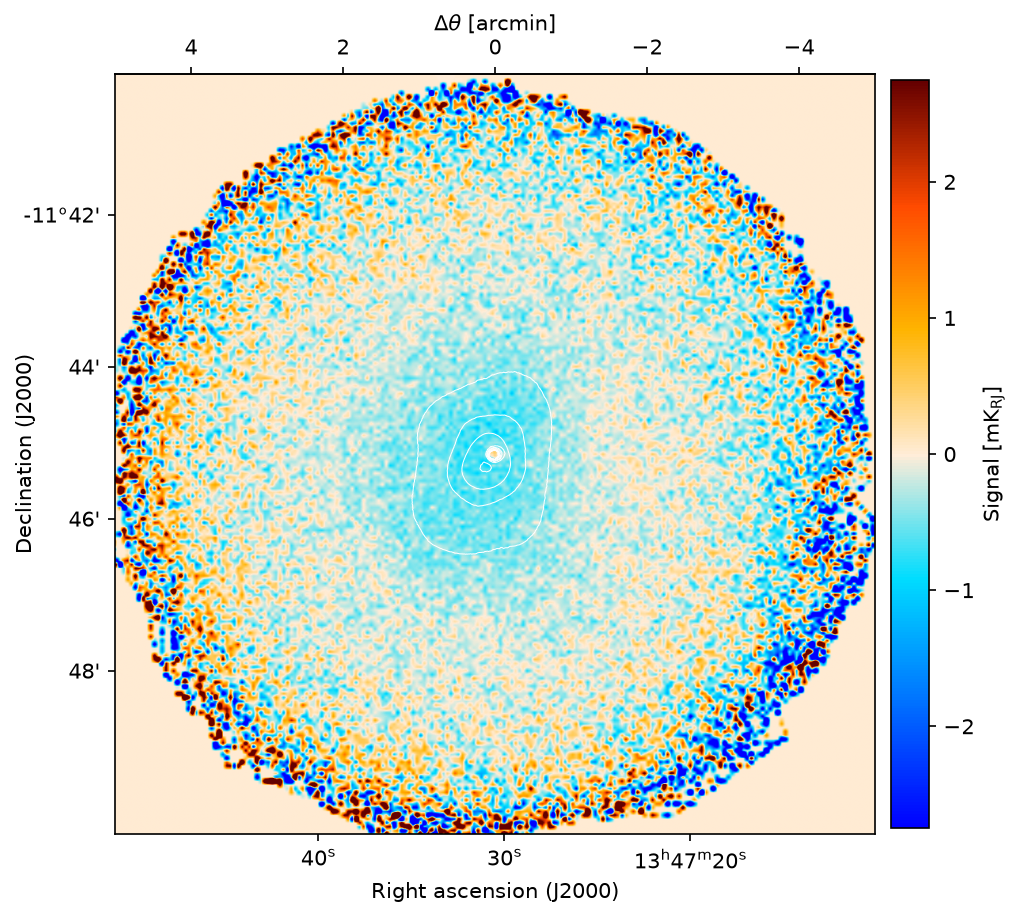}}
  \caption{Difference between the beam-convolved posterior mean of the \texttt{maria-nifty} map model and the \texttt{minkasi} map, shown for the full $10\arcmin \times 10\arcmin$ FOV of \cref{fig:map-comparison}. The white contours indicate the SZ decrement and the compact source of the posterior mean. The negative difference of about $-1\,\mathrm{mK}_{\mathrm{RJ}}$ across the cluster reflects the deeper decrement recovered by our reconstruction. Towards the edge of the scanned region, the difference is dominated by the strongly increasing noise of the \texttt{minkasi} map.}
  \label{fig:map-difference}
\end{figure}

The map model is defined on a grid that is considerably finer than the angular resolution of the instrument. To show the reconstruction at the resolution at which it is constrained by the data, \cref{fig:map-smoothed} depicts the posterior mean of \cref{fig:map-reconstruction} convolved with the beam of MUSTANG-2. To compare the large-scale signal of the two reconstructions, \cref{fig:map-difference} shows the difference between the beam-convolved posterior mean and the \texttt{minkasi} map of \cref{fig:map-comparison}. Within the cluster, the difference is negative throughout, so our reconstruction recovers a deeper SZ decrement over the whole extent of the cluster rather than only in individual pixels. Outside the cluster, the difference shows no coherent large-scale structure across the FOV and is dominated by the small-scale noise of the \texttt{minkasi} map, which increases strongly towards the edge of the scanned region.

\section{Comparison to ALMA} \label{app:alma}

\begin{figure*}[t]
  \centering
  \includegraphics[width=\textwidth]{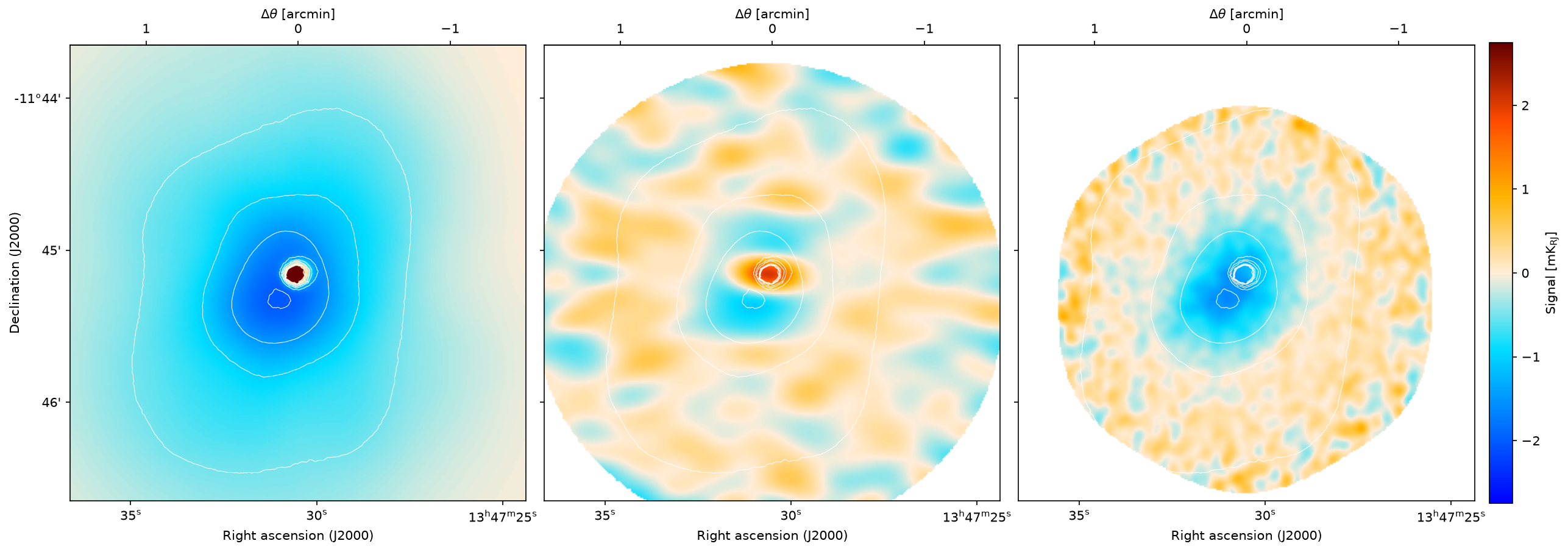}
  \caption{Comparison of our reconstruction of RXJ1347 to ALMA observations at 92~GHz \citep{kitayama2016}. \textit{Left:} the posterior mean of the \texttt{maria-nifty} map model, as in \cref{fig:map-reconstruction}. \textit{Middle:} the dirty map of the ALMA 7-m array before the subtraction of the central AGN (Fig.~1 of \citealt{kitayama2016}). \textit{Right:} the deconvolved ALMA map of the combined 7-m and 12-m arrays after the subtraction of the AGN, smoothed to a beam of $5\arcsec$ FWHM (Fig.~4 of \citealt{kitayama2016}). The white contours indicate the reconstructed SZ decrement and compact source of our posterior mean, and all panels share the colour scale on the right and show the inner $3\arcmin \times 3\arcmin$ around the cluster centre.}
  \label{fig:map-alma}
  \vspace{1cm}
\end{figure*}

RXJ1347 has also been observed with ALMA in Band~3 at a central frequency of 92~GHz, using a seven-pointing hexagonal mosaic of both the 7-m and the 12-m arrays \citep{kitayama2016}. \Cref{fig:map-alma} compares the posterior mean of our map model to two of the ALMA images presented in that work: the dirty map of the 7-m array before the subtraction of the central AGN and the deconvolved map of the combined arrays after its subtraction, smoothed to a beam of $5\arcsec$ FWHM. Given the ALMA interferometer's spacing and dish sizes, the 12-meter array of ALMA is insensitive to angular scales greater than $\sim 40\arcsec$, so that its images only contain the core of the cluster and lack the extended SZ decrement recovered in our reconstruction. Within the core, however, the maps are in good agreement. The compact source of our second map component coincides with the AGN in the 7-m map, whose elongation in the east-west direction is caused by the synthesized beam of $20.5\arcsec \times 11.1\arcsec$. In both our reconstruction and the deconvolved ALMA map, the deepest part of the decrement is located south-east of the AGN, corresponding to the south-east peak identified by \citet{kitayama2016}.

\section{Prior and posterior samples} \label{app:additional-plots}

\begin{figure*}
\centering
   \includegraphics[width=17cm]{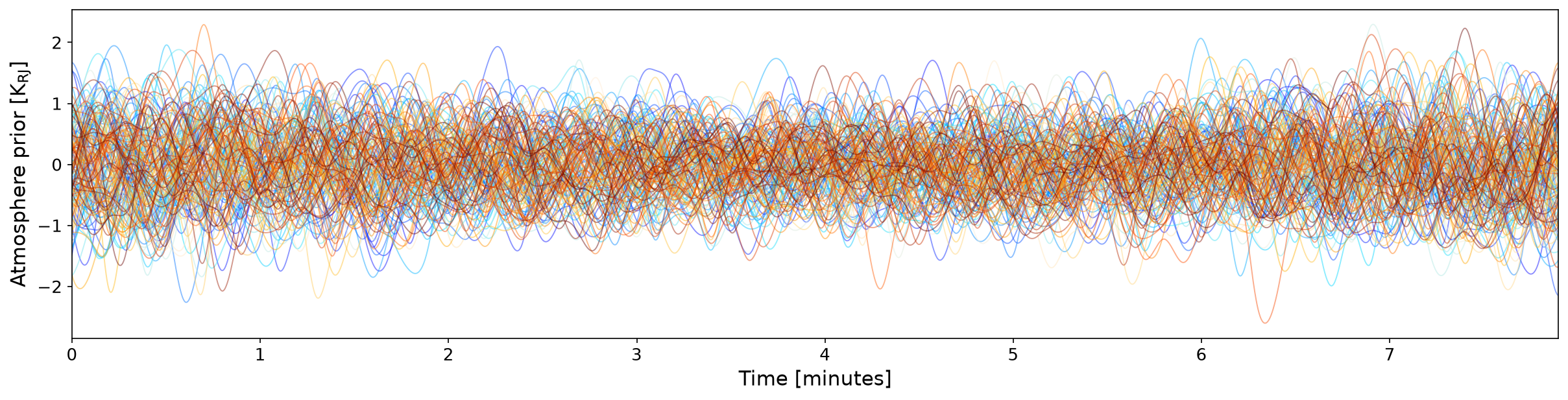}
     \caption{One prior sample of the atmospheric model for each detector of a single scan of RXJ1347, with the detectors coloured according to their index as in \cref{fig:data}. The curves are drawn from the cubic spline model of \cref{eq:bspline_el} and vary smoothly over the eight minutes of the scan, with amplitudes of the order of one Kelvin, which is the same order as the atmospheric emission in the observed TODs.}
     \label{fig:tod-atmosphere-prior}
\end{figure*}

\begin{figure*}
\centering
   \includegraphics[width=17cm]{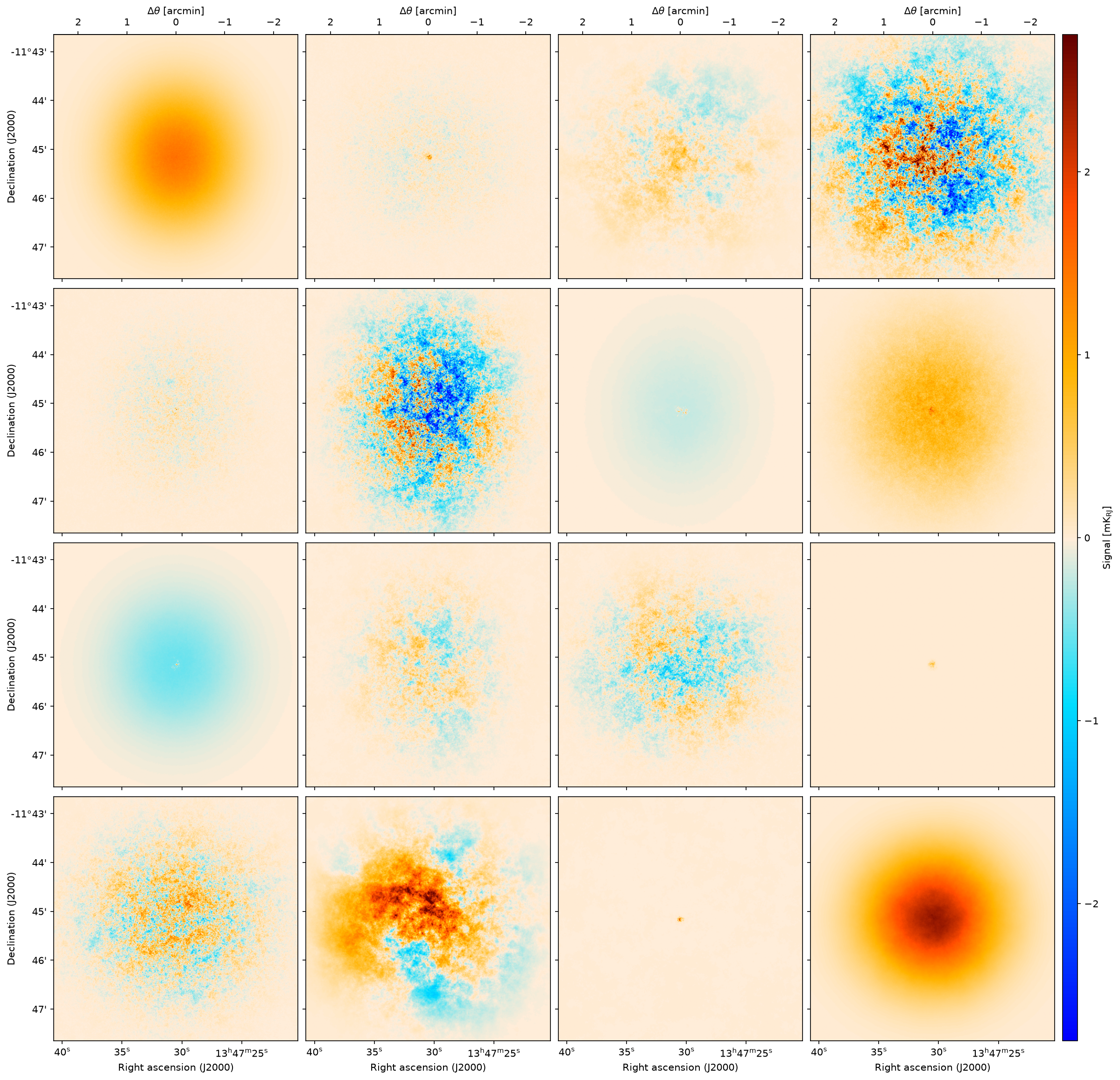}
     \caption{Sixteen prior samples of the two-component map model used for the reconstruction of RXJ1347. Each panel shows the sum of both components, which are drawn from the CFM of \texttt{NIFTy} with the hyperparameters listed in \cref{tab:params}. The samples feature correlated structures on a range of spatial scales, both positive and negative, and fall off towards the edges of the FOV due to the Gaussian window applied to the output of the CFM.}
     \label{fig:map_prior}
\end{figure*}

\begin{figure*}
\centering
   \includegraphics[width=17cm]{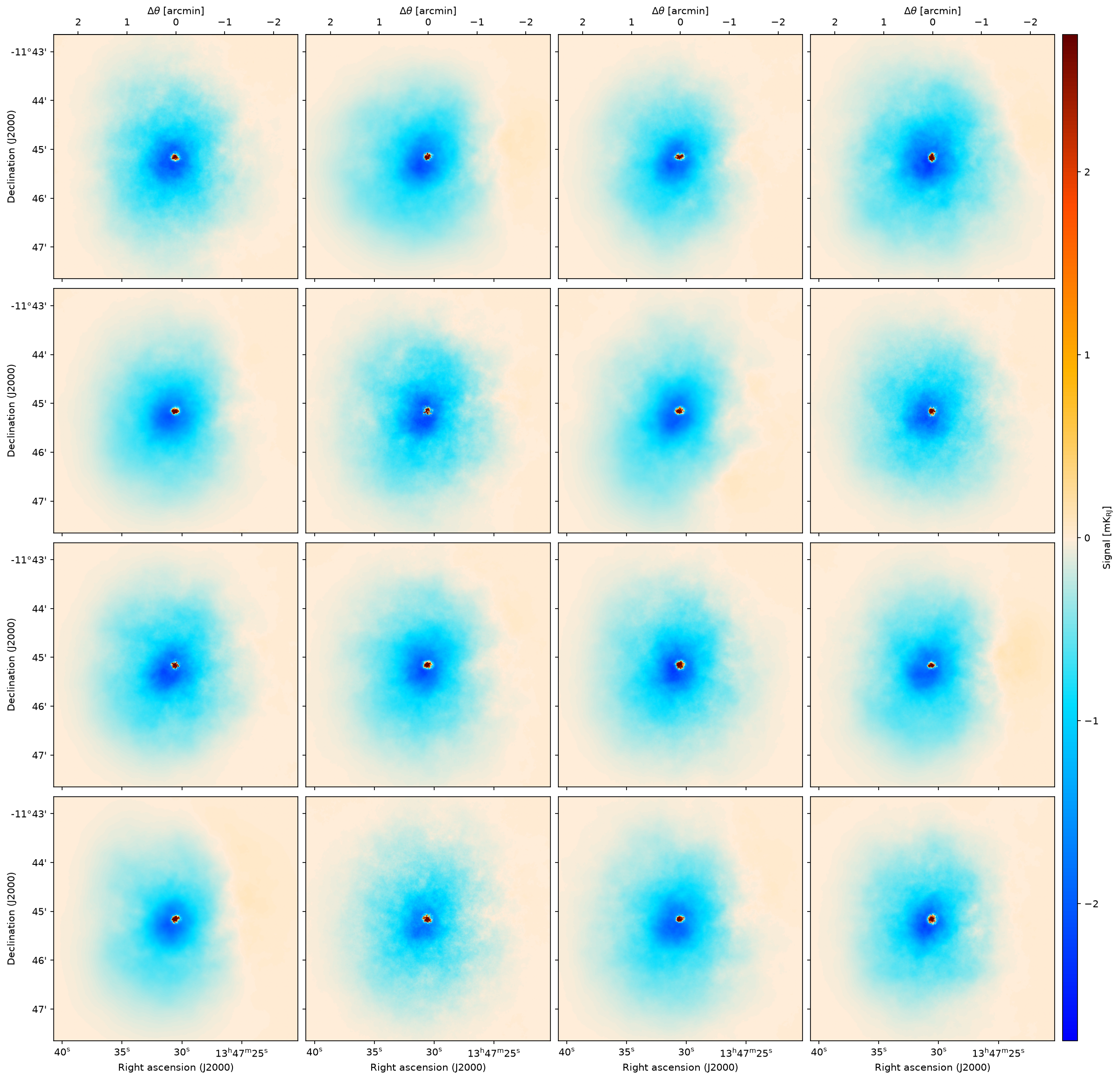}
     \caption{The eight posterior samples of the map model that underlie the reconstruction shown in \cref{fig:map-reconstruction}. In contrast to the prior samples of \cref{fig:map_prior}, all of them show the SZ decrement of the cluster and the compact source at its centre. They differ mainly in the outer parts of the map, where the signal is faint and therefore less constrained by the data, which is reflected in the relative uncertainty shown in \cref{fig:map-reconstruction}.}
     \label{fig:map-posterior}
\end{figure*}

This section collects additional figures of the RXJ1347 reconstruction that are not shown in the main text. \Cref{fig:tod-atmosphere-prior} depicts prior samples of the atmospheric model in data space for one of the ten scans. \Cref{fig:map_prior} shows prior samples of the two-component map model, illustrating which structures the CFM supports before any data are taken into account, while \cref{fig:map-posterior} shows the corresponding posterior samples that underlie the reconstruction of \cref{fig:map-reconstruction}.

\end{appendix}

\end{document}